\documentclass[11pt]{article}
\usepackage[margin=0.7in]{geometry}
\usepackage{amsmath}
\usepackage{amssymb,amsthm}
\usepackage{algorithm,booktabs}
\usepackage{algpseudocode}
\usepackage{graphicx}
\usepackage{url}
\usepackage{dsfont,mathtools}
\usepackage[usenames,dvipsnames]{xcolor}
\definecolor{darkgreen}{rgb}{0,0.66,0}
\usepackage[caption=false,font=footnotesize]{subfig}
\usepackage{hyperref}
\usepackage{tikz}
\usepackage{pgfplots}
\usetikzlibrary{3d,calc,patterns,shadings}

\renewcommand{\vec}[1]{\mathbf{#1}}
\renewcommand{\vec}[1]{\boldsymbol{#1}}

\newcommand{\eps}{\varepsilon}
\newcommand{\vc}[1]{\boldsymbol{#1}}

\newcommand{\cC}{\mathcal{C}}
\newcommand{\cD}{\mathcal{D}}

\newcommand{\cH}{\mathcal{H}}

\newcommand{\cS}{\mathcal{S}}

\newcommand{\cW}{\mathcal{W}}

\newcommand{\R}{\mathbb{R}}

\newcommand{\tO}{\tilde{O}}

\newcommand{\ip}[2]{\left\langle{#1},{#2}\right\rangle}
\newcommand{\rhoq}{\rho_{\mathrm{q}}}

\newcommand{\rhou}{\rho_{\mathrm{u}}}
\newcommand{\alphaq}{\alpha_{\mathrm{q}}}
\newcommand{\alphau}{\alpha_{\mathrm{u}}}
\newcommand{\alphal}{\alpha_{\mathrm{low}}}
\newcommand{\alphah}{\alpha_{\mathrm{high}}}

\newtheorem{theorem}{Theorem}
\newtheorem{corollary}{Corollary}
\newtheorem{definition}{Definition}
\newtheorem{conjecture}{Conjecture}

\title{\textbf{Tradeoffs for nearest neighbors on the sphere}}

\author{Thijs Laarhoven\thanks{Eindhoven University of Technology, Eindhoven, The Netherlands. E-mail: \texttt{mail@thijs.com}}}

\begin{document}

\maketitle


\begin{abstract}
We consider tradeoffs between the query and update complexities for the (approximate) nearest neighbor problem on the sphere, extending the spherical filters recently introduced by [Becker--Ducas--Gama--Laarhoven, SODA'16] to sparse regimes and generalizing the scheme and analysis to account for different tradeoffs. In a nutshell, for the sparse regime the tradeoff between the query complexity $n^{\rhoq}$ and update complexity $n^{\rhou}$ for data sets of size $n$ can be summarized by the following equation in terms of the approximation factor $c$ and the exponents $\rhoq$ and $\rhou$:
\begin{align*}
c^2 \sqrt{\rhoq} + (c^2 - 1) \sqrt{\rhou} = \sqrt{2c^2 - 1}.
\end{align*}

For small $c = 1 + \eps$, minimizing the time for updates leads to a linear space complexity at the cost of a query time complexity of approximately $n^{1 - 4 \eps^2}$. Balancing the query and update costs leads to optimal complexities of $n^{1/(2c^2 - 1)}$, matching lower bounds from [Andoni--Razenshteyn, 2015] and [Dubiner, IEEE Trans.\ Inf.\ Theory 2010] and matching the asymptotic complexities previously obtained by [Andoni--Razenshteyn, STOC'15] and [Andoni--Indyk--Laarhoven--Razenshteyn--Schmidt, NIPS'15]. A subpolynomial query time complexity $n^{o(1)}$ can be achieved at the cost of a space complexity of the order $n^{1/(4\eps^2)}$, matching the lower bound $n^{\Omega(1/\eps^2)}$ of [Andoni--Indyk--P\v{a}tra\c{s}cu, FOCS'06] and [Panigrahy--Talwar--Wieder, FOCS'10] and improving upon results of [Indyk--Motwani, STOC'98] and [Kushilevitz--Ostrovsky--Rabani, STOC'98] with a considerably smaller leading constant in the exponent.

For large $c$, minimizing the update complexity results in a query complexity of $n^{2/c^2 + O(1/c^4)}$, improving upon the related asymptotic exponent for large $c$ of [Kapralov, PODS'15] by a factor $2$, and matching the lower bound $n^{\Omega(1/c^2)}$ of [Panigrahy--Talwar--Wieder, FOCS'08]. Balancing the costs leads to optimal complexities of the order $n^{1/(2c^2 - 1)}$, while a minimum query time complexity can be achieved with update and space complexities of approximately $n^{2/c^2 + O(1/c^4)}$ and $n^{1 + 2/c^2 + O(1/c^4)}$, also improving upon the previous best exponents of Kapralov by a factor $2$ for large $n$ and $c$.

For the regime where $n$ is exponential in the dimension, we obtain further improvements compared to results obtained with locality-sensitive hashing. We provide explicit expressions for the query and update complexities in terms of the approximation factor $c$ and the chosen tradeoff, and we derive asymptotic results for the case of the highest possible density for random data sets.
\end{abstract}


\section{Introduction}

\paragraph{Approximate nearest neighbors (ANN).} A central computational problem in many areas of research, such as machine learning, coding theory, pattern recognition, data compression, and cryptanalysis~\cite{bishop06, dubiner10, duda00, herold15, laarhoven15lsh, may15, shakhnarovich05}, is the \textit{nearest neighbor problem}: given a $d$-dimensional data set $\cD \subset \R^d$ of cardinality $n$, design a data structure and preprocess $\cD$ in a way that, when later given a query vector $\vc{q} \in \R^d$, we can quickly find a nearest vector to $\vc{q}$ in $\cD$. A common relaxation of this problem is the \textit{approximate nearest neighbor problem (ANN)}: given that the nearest neighbor in $\cD$ lies at distance at most $r$ from $\vc{q}$, design an efficient algorithm that finds an element $\vc{p} \in \cD$ at distance at most $c \cdot r$ from $\vc{q}$, for a given approximation factor $c > 1$. We will consider the case where $d$ scales with $n$; for fixed $d$ it is well-known that one can answer queries in time$n^{\rho}$ with $\rho = o(1)$ with only a polynomial increase in the space complexity~\cite{arya94}.

\paragraph{ANN on the sphere.} Depending on the notion of distance, different solutions have been proposed in the literature (e.g.\ \cite{alt01, datar04, gionis99, leskovec14, wang14}). In this work we will restrict out attention to the \textit{angular distance}, where two vectors are considered nearby iff their common angle is small~\cite{andoni15cp, charikar02, laarhoven15lsh, schmidt14, sundaram13}. This equivalently corresponds to spherical ANN under the $\ell_2$-norm, where the entire data set is assumed to lie on the Euclidean unit sphere. Recent work of Andoni and Razenshteyn~\cite{andoni15} showed how to reduce ANN in the entire Euclidean space to ANN on the sphere, which further motivates why finding optimal solutions for the spherical case is relevant. For spherical, low-density settings ($n = 2^{o(d)}$), we will further focus on the \textit{random} setting described in~\cite{andoni15}, where \textit{nearby} corresponds to a Euclidean distance of $r = \frac{1}{c}\sqrt{2}$ on the unit sphere (i.e.\ an angle $\theta = \arccos(1 - 1/c^2)$) and \textit{far away} corresponds to a distance $c \cdot r = \sqrt{2}$ (angle $\psi = \frac{1}{2}\pi$). 

\paragraph{Spherical locality-sensitive hashing.} A well-known method for solving ANN in high dimensions is \textit{locality-sensitive hashing (LSH)}~\cite{indyk98}. Using locality-sensitive hash functions, with the property that nearby vectors are more likely to be mapped to the same output value than distant pairs of vectors, one builds several hash tables with buckets containing sets of vectors with the same hash value. To answer a query $\vc{q}$, one computes $\vc{q}$'s hash values and checks the corresponding buckets in each of the hash tables for potential near neighbors. For spherical ANN in the random setting, two recent works~\cite{andoni15, andoni15cp} have shown how to solve ANN with query time $\tO(n^{\rho})$ and space $\tO(n^{1 + \rho})$ with $\rho = \frac{1}{2c^2 - 1} + o(1)$ where $o(1) \to 0$ as $n \to \infty$. For large $c$ and $n$, this improved upon e.g.\ hyperplane LSH~\cite{charikar02} and Euclidean LSH~\cite{andoni06}. Within the class of LSH algorithms, these results are known to be essentially optimal~\cite{andoni15lb, dubiner10}.

\paragraph{Spherical locality-sensitive filters.} Recently Becker--Ducas--Gama--Laarhoven~\cite{becker15lsf} introduced spherical filters, which map the data set $\cD$ to a subset $\cD' \subseteq \cD$ consisting of all points lying in a certain spherical cap. Filtering could be considered a relaxation of locality-sensitive hashing: for LSH a hash function is required to \textit{partition} the space in regions, while for LSF this is not necessarily the case. Similar filtering constructions were previously proposed in~\cite{dubiner10, may15}. For dense data sets of size $n = 2^{\Theta(d)}$, the approach of~\cite{becker15lsf} led to a query exponent $\rho < 1/(2c^2 - 1)$ for the random setting.

\paragraph{Asymmetric ANN.} The exponents $\rho$ described so far are all for \textit{balanced} or \textit{symmetric} ANN: both the time to answer a query and the time to insert/delete vectors from the data structure are then equal to $\tO(n^{\rho})$, and the time complexity for preprocessing the data and the total space complexity are both equal to $\tO(n^{1 + \rho})$. Depending on the application however, it may be desirable to obtain a different tradeoff between these costs. In some cases it may be beneficial to use even more space and more time for the preprocessing, so that queries can be answered even faster. In other cases, memory constraints might rule out the use of balanced parameters, in which case one has to settle for a lower space and update complexity and it would be interesting to know the best time complexity that can be achieved for a given space complexity. Finding optimal tradeoffs between the different costs of ANN is therefore essential for achieving the best performance in different contexts.

\paragraph{Smooth tradeoffs for asymmetric ANN.} Various works have analyzed tradeoffs for ANN, among others using multi-probing in LSH to reduce the memory complexity at the cost of a higher query complexity~\cite{andoni15cp, arya09, lv07, panigrahy06}. However, most existing techniques either describe one particular tradeoff between the costs, or do not offer provable asymptotic bounds on the query and update exponent as the parameters increase. Recently Kapralov~\cite{kapralov15} showed how to obtain smooth and provable asymptotic tradeoffs for Euclidean ANN, but as the exponents for the balanced setting are a factor $2$ above the lower bound $\rho \geq 1/(2c^2 - 1)$ for large $c$, it may be possible to improve upon these techniques not only for symmetric but also for asymmetric ANN.

\subsection{Contributions.} In this work we extend the symmetric ANN technique of~\cite{becker15lsf} for dense regimes to asymmetric ANN on the sphere for both sparse and dense regimes, showing how to obtain smooth and significantly improved tradeoffs between the query and update complexities compared to e.g.~\cite{kapralov15, panigrahy06} in both the small $c$ and large $c$ regimes. For sparse settings, the tradeoff between the query complexity $n^{\rhoq}$ and the update complexity $n^{\rhou}$ can essentially be summarized by the non-negative solution pairs $(\rhou, \rhoq) \in \R^2$ to the following equation, which can be expressed either in terms of $\theta$ (left) or $c$ (right) by substituting $\cos \theta = 1 - 1/c^2$.
\begin{align}
\sqrt{\rhoq} + (\cos \theta) \sqrt{\rhou} = \sin \theta, \qquad \qquad c^2 \sqrt{\rhoq} + (c^2 - 1) \sqrt{\rhou} = \sqrt{2 c^2 - 1}. \label{eq:main}
\end{align}
The space complexity for the preprocessed data, as well as the time for preprocessing the data, are both $\tO(n^{1 + \rhou})$. The resulting tradeoffs for the random case for small and large $c$ are illustrated in Table~\ref{tab:tradeoff} and Figure~\ref{fig:tradeoff}, and can be derived from~\eqref{eq:main} by substituting $\rhou = 0$ (minimize the space), $\rhoq = \rhou$ (balanced ANN), or $\rhoq = 0$ (minimize the time), and computing Taylor expansions around $c \in \{1, \infty\}$.


\paragraph{Small approximation factors.} In the regime of small $c = 1 + \eps$, as described in Table~\ref{tab:tradeoff} we obtain an update complexity $n^{o(1)}$ and space complexity $n^{1 + o(1)}$ with query complexity $n^{1 - 4 \eps^2 + O(\eps^3)}$. This improves upon results of Kapralov, where a sublinear query complexity and a quasi-linear space complexity could only be achieved for approximation factors $c > 1.73$~\cite{kapralov15}. Balancing the complexities leads to asymptotic exponents $\rhoq = \rhou = 1/(2c^2 - 1)$, which means that both exponents scale as $1 - O(\eps)$ for small $c > 1$. These exponents match the asymptotic complexities previously obtained by~\cite{andoni15, andoni15cp} and the lower bounds from~\cite{andoni15lb, dubiner10}. A query complexity $n^{o(1)}$ can be achieved for arbitrary $c$ with an update complexity $n^{4/\eps^2 + O(1/\eps)}$, matching the asymptotic lower bounds of~\cite{andoni06b, panigrahy10}\footnote{The constant $1/4$ in the exponent even matches the lower bound for single-probe schemes of~\cite[Theorem 1.5]{panigrahy10}.} and the constructions of~\cite{indyk98, kushilevitz00} with a smaller leading constant in the exponent\footnote{Explicit leading constants are not stated in~\cite{indyk98, kushilevitz00} but appear to be $9$ and $9/2$ respectively, compared to our $1/4$.}. This also improves upon~\cite{kapralov15}, achieving a query complexity $n^{o(1)}$ only for $c > 1.73$.

\paragraph{Large approximation factors.} For large $c$, both $\rhoq$ and $\rhou$ are proportional to $1/c^2$, with leading constants $1/2$ in the balanced regime, and leading constant $2$ if the other complexity is minimized. This improves upon results from~\cite{kapralov15}, whose leading constants are a factor $2$ higher in all cases, and matches the lower bound on the space complexity of $n^{\Omega(1/c^2)}$ of~\cite{panigrahy08} for query complexities $n^{o(1)}$. 

\paragraph{High-density regime.} Finally, for data sets of size $n = 2^{\Theta(d)}$, we obtain improved tradeoffs between the query and update complexities compared to results obtained using locality-sensitive hashing, even for balanced settings. We show that also for this harder problem we obtain query exponents less than $1$ regardless of the tradeoff, while a query exponent $0$ is impossible to achieve with our methods.

\subsection{Outline.} In Section~\ref{sec:prelim} we describe preliminary notation and results regarding spherical filters. Section~\ref{sec:sparse} describes asymptotic tradeoffs for the sparse regime of $n = 2^{o(d)}$, and Section~\ref{sec:dense} then discusses the application of these techniques to the dense regime of $n = 2^{\Theta(d)}$. Section~\ref{sec:discussion} concludes with a discussion on extending our methods to slightly different problems, and open problems for future work. 


\begin{table*}[!t]
\centering
\renewcommand{\arraystretch}{1.1}
\begin{tabular}{p{3.13cm}p{4.37cm}p{4.01cm}p{4.04cm}} \toprule
 & \textbf{General expressions} & \textbf{Small $c = 1 + \eps$} & \textbf{Large $c \to \infty$} \\ \midrule 
$\left. \begin{matrix*}[l] 
\text{\bfseries Minimize space} \\ 
\ \ (\beta = \cos \theta)
\end{matrix*}\right.$ & $\left. \begin{matrix*}[l] 
\mathbf{\color{red}\rhoq = (2c^2 - 1)/c^4} \\ 
\mathbf{\color{red}\rhou = 0}
\end{matrix*}\right.$ & $\left. \begin{matrix*}[l] 
\rhoq = 1 - 4 \eps^2 + O(\eps^3) \\ 
\rhou = 0
\end{matrix*}\right.$ & $\left. \begin{matrix*}[l] 
\rhoq = 2/c^2 + O(1/c^4) \\ 
\rhou = 0
\end{matrix*}\right.$ \\ \midrule
$\left. \begin{matrix*}[l] 
\text{\bfseries Balance costs} \\ 
\ \ (\beta = 1)
\end{matrix*}\right.$ & $\left. \begin{matrix*}[l] 
\mathbf{\color{darkgreen}\rhoq = 1/(2c^2 - 1)} \\ 
\mathbf{\color{darkgreen}\rhou = 1/(2c^2 - 1)}
\end{matrix*}\right.$ & $\left. \begin{matrix*}[l] 
\rhoq = 1 - 4 \eps + O(\eps^2) \\ 
\rhou = 1 - 4 \eps + O(\eps^2)
\end{matrix*}\right.$ & $\left. \begin{matrix*}[l] 
\rhoq = 1/(2c^2) + O(1/c^4) \\ 
\rhou = 1/(2c^2) + O(1/c^4)
\end{matrix*}\right.$ \\ \midrule
$\left. \begin{matrix*}[l] 
\text{\bfseries Minimize time} \\ 
\ \ (\beta = 1 / \cos \theta)
\end{matrix*}\right.$ & $\left. \begin{matrix*}[l] 
\mathbf{\color{blue}\rhoq = 0} \\ 
\mathbf{\color{blue}\rhou = (2 c^2 - 1)/(c^2 - 1)^2}
\end{matrix*}\right.$ & $\left. \begin{matrix*}[l] 
\rhoq = 0 \\ 
\rhou = 1/(4 \eps^2) + O(1/\eps)
\end{matrix*}\right.$ & $\left. \begin{matrix*}[l] 
\rhoq = 0 \\ 
\rhou = 2/c^2 + O(1/c^4)
\end{matrix*}\right.$ \\ \bottomrule
\end{tabular}
\caption{The extreme points of our asymptotic tradeoffs. Answering a query takes time $\tO(n^{\rhoq})$, updates take $\tO(n^{\rhou})$ operations, and the space/preprocessing complexities are $\tO(n^{1 + \rhou})$. Lower order terms which tend to $0$ as $d, n \to \infty$ are omitted for clarity. The colors match those used in Figure~\ref{fig:tradeoff}. \label{tab:tradeoff}}
\end{table*}


\section{Preliminaries}
\label{sec:prelim}

\subsection{Subsets of the unit sphere.} We first recall some preliminary notation and results on geometric objects on the unit sphere, similar to~\cite{becker15lsf}. Let $\mu$ denote the canonical Lebesgue measure over $\R^d$, and let us write $\ip{\cdot}{\cdot}$ for the standard Euclidean inner product. We denote the unit sphere in $\R^d$ by $\cS^{d-1} = \{\vec x \in \R^d: \|\vec x\| = 1\}$ and half-spaces by $\cH_{\vec u, \alpha} := \{\vec x \in \R^d: \ip{\vec u}{\vec x} \geq \alpha\}$. For constants $\alpha, \alpha_1, \alpha_2 \in (0,1)$ and vectors $\vec{u}, \vec u_1, \vec u_2 \in \cS^{d-1}$ we denote spherical caps and wedges by $\cC_{\vec u, \alpha} := \cS^{d-1} \cap \cH_{\vec u, \alpha}$ and $\cW_{\vec u_1, \alpha_1, \vec u_2, \alpha_2} := \cS^{d-1} \cap \cH_{\vec u_1,\alpha_1} \cap \cH_{\vec u_2, \alpha_2}$.

For analyzing the performance of spherical filters, we would like to know the volumes of these objects in high dimensions. The following asymptotic estimates can be found in \cite{becker15lsf, micciancio10b}, where $\gamma = \gamma(\alpha_1, \alpha_2, \theta)$ satisfies $\gamma^2 = (\alpha_1^2 + \alpha_2^2 - 2 \alpha_1 \alpha_2 \cos \theta) / \sin^2 \theta$ and $\theta$ denotes the angle $\phi(\vec u_1, \vec u_2) := \arccos \ip{\vc u_1}{\vc u_2}$ between $\vc{u}_1, \vc{u}_2$. 
\begin{align}
\cC(\alpha) := \frac{\mu(\cC_{\vec u,\alpha})}{\mu(\cS^{d-1})} = d^{\Theta(1)} \left(\sqrt{1 - \alpha^2}\right)^d, \quad  \cW(\alpha_1,\alpha_2,\theta) := \frac{\mu(\cW_{\vec u_1, \alpha_1, \vec u_2, \alpha_2})}{\mu(\cS^{d-1})} = d^{\Theta(1)} \left(\sqrt{1 - \gamma^2}\right)^d.
\end{align}

\subsection{Symmetric spherical filters in the dense regime.} We now continue with a brief description of the algorithm of Becker--Ducas--Gama--Laarhoven~\cite{becker15lsf} for solving dense ANN on the sphere. 


\paragraph{Initialization.} Let $m = \Theta(\log d)$ and suppose that $m | d$. We partition the $d$ coordinates into $m$ blocks of size $d / m$, and for each of these $m$ blocks of coordinates we randomly sample $t^{1/m}$ code words from $\cS^{d/m - 1}$. This results in $m$ \textit{subcodes} $C_1, \dots, C_m \subset \cS^{d/m - 1}$. Combining one code word from each subcode, we obtain $(t^{1/m})^m = t$ different vectors $\frac{1}{\sqrt{m}} (\vc{c}_1, \dots, \vc{c}_m) \in \cS^{d-1}$ with $\vc{c}_i \in C_i$. We denote the resulting set of vectors by the \textit{code} $C$. The ideas behind this construction are that (1) this code $C$ behaves as a set of $t$ random unit vectors in $\cS^{d-1}$, where the difference with a completely random code is negligible for large parameters~\cite[Theorem 5.1]{becker15lsf}; and (2) the additional structure hidden in $C$ allows us to \textit{decode} faster than with a linear search. The parameter $t$ will be specified later.

\paragraph{Preprocessing.} Next, given $\cD \subset \cS^{d-1}$, we consider each point $\vc{p} \in \cD$ one by one and compute its relevant filters $\text{Update}(\vc{p}) := \{\vc{c} \in C: \ip{\vc{p}}{\vc{c}} \geq \alpha\}$. Naively finding these filters by a linear search over all filters would cost time $\tO(t)$, but as described in~\cite{becker15lsf} this can be done in time $O(|\text{Update}(\vc{p}_i)|)$ due to the hidden additional structure in the code\footnote{Note that the overhead of the enumeration-based decoding algorithm~\cite[Algorithm~1]{becker15lsf} mainly consists of computing and sorting all blockwise inner products $\ip{\vec p_i}{\vec c_i}$, which can be done in time $\tO(t^{1/m}) = n^{o(1)}$.}. Finally, we store all vectors in the respective filter buckets $B_1, \dots, B_t$, where $\vc{p}$ is stored in $B_j$ iff $\vc{c}_j \in \text{Update}(\vc{p})$. The parameter $\alpha$ will be specified later.

\paragraph{Answering a query.} To find neighbors for a query vector $\vc{q}$, we compute its relevant filters $\text{Query}(\vc{q}) := \{\vc{c} \in C: \ip{\vc{q}}{\vc{c}} \geq \alpha\}$ in time proportional to the size of this set. Then, we visit all these buckets in our data structure, and compare $\vc{q}$ to all vectors $\vc{p}$ in these buckets. The cost of this step is proportional to the number of vectors colliding with $\vc{q}$ in these filters, and the success probability of answering a query depends on the probability that two nearby vectors are found due to a collision.

\paragraph{Updating the data structure (optional).} In certain applications, it may further be important that one can efficiently update the data structure when $\cD$ is changed. Inserting or removing a vector $\vc{p}$ from the buckets is done by computing $\text{Update}(\vc{p})$ in time proportional to $|\text{Update}(\vc{p})|$, and inserting/removing the vector from the corresponding buckets. Note that by e.g.\ keeping buckets sorted in lexicographic order, updates in one bucket can be done in time $\tO(\log n) = d^{O(1)}$.

\paragraph{Correctness.} To prove that this filtering construction works for certain parameters $\alpha$ and $t$, two properties are crucial: the code $C$ needs to be efficiently decodable, and $C$ must be sufficiently smooth on $\cS^{d-1}$ in the sense that collision probabilities are essentially equal to those of uniformly random codes $C \subset \cS^{d-1}$. These two properties were proved in~\cite[Lemma 5.1 and Theorem 5.1]{becker15lsf} respectively.


\section{Asymmetric spherical filters for sparse regimes} 
\label{sec:sparse}

To convert the spherical filter construction described above to the low-density regime, we need to make sure that the overhead remains negligible in $n$. Note that costs $t^{1/m} = t^{1/\log d}$ are considered $n^{o(1)}$ in~\cite{becker15lsf} as $t = 2^{\Theta(d)}$ and $n = 2^{\Theta(d)}$. In the sparse setting of $n = 2^{\Theta(d / \log d)}$, this may no longer be the case\footnote{For even sparser data sets, we can always first apply a dimension reduction using the Johnson-Lindenstrauss transform~\cite{johnson84} to transform the points to $d'$-dimensional vectors with $n = 2^{\Omega(d' / \log d')}$, and without significantly distorting inter-point distances.}. To overcome this potential issue, we set $m = O(\log^2 d)$, so that $t^{1/m} = n^{o(1)}$ even if $t = 2^{\Theta(d)}$. Increasing $m$ means that the code $C$ becomes less smooth, but a detailed inspection of the proof of~\cite[Thm.~5.1]{becker15lsf} shows that also for $m = \log^{O(1)} d$ the code is sufficiently smooth on $\cS^{d-1}$.

To allow for tradeoffs between the query/update complexities, we introduce two parameters $\alphaq$ and $\alphau$ for querying and updating the database. This means that we redefine $\text{Query}(\vc{q}) := \{\vc{c} \in C: \ip{\vc{q}}{\vc{c}} \geq \alphaq\}$ and $\text{Update}(\vc{p}) := \{\vc{c} \in C: \ip{\vc{p}}{\vc{c}} \geq \alphau\}$ where $\alphaq, \alphau \in (0,1)$ are to be chosen later. Smaller parameters mean that more filters are contained in these sets, so intuitively $\alphaq < \alphau$ means more time is spent on queries, while $\alphaq > \alphau$ means more time is spent on updates and less time on queries.

\subsection{Random sparse instances.}\label{sec:sparse-random} For studying the sparse regime of $n = 2^{o(d)}$, we will consider the \textit{random} model of~\cite{andoni15, andoni15cp} defined as follows.

\begin{definition}[Random $\theta$-ANN in the sparse regime]
Given an angle $\theta \in (0, \frac{1}{2} \pi)$, a query $\vc{q} \in \R^d$, and a data set $\cD$ of size $n = 2^{o(d)}$, the $\theta$-ANN problem is defined as the problem of either finding a point $\vc{p} \in \cD$ with $\phi(\vc{p}, \vc{q}) \leq \frac{1}{2} \pi$, or concluding that w.h.p.\ no vector $\vc{p} \in \cD$ exists with $\phi(\vc{p}, \vc{q}) \leq \theta$.
\end{definition}

Note that for angles $\psi = \frac{1}{2} \pi - \delta$ where $\delta > 0$ is fixed independently of $d$ and $n$, a random point $\vc{u}$ on the sphere \textit{covers} a fraction $(1 - O(\delta^2))^d = 2^{-\Theta(d)}$ of the sphere with points at angle at most $\psi$ from $\vc{u}$. Together, $n = 2^{o(d)}$ points therefore cover a fraction of at most $t \cdot 2^{-\Theta(d)} = 2^{-\Theta(d)}$ of the sphere. For a query $\vc{q}$ sampled uniformly at random from the sphere, with high probability it is far away from all $n$ points, i.e.\ at angle at least $\psi$ from $\cD$. In other words, we expect that there is a significant gap between the angle with the (planted) nearest neighbor, and the angle with all other vectors, in which case solving ANN with small approximation factor is actually equivalent to solving the exact NN problem.

If we were to define the notion of ``far away'' as being at some angle $\psi < \frac{1}{2} \pi$ from $\vc{q}$, and we somehow expect that a significant part of the data set lies at angle at most $\psi$ from $\vec q$, then the data set and queries are apparently concentrated on one part of the sphere and their distribution is not spherically symmetric. If this is indeed the case, then this knowledge may be exploited using data-dependent ANN techniques~\cite{wang14}, and such techniques may be preferred over data-independent filters.

\subsection{Main result.} Before describing the analysis that leads to the optimized parameter choices, we state the main result for the random, sparse setting described above, in terms of the nearby angle $\theta$.
	\begin{theorem} \label{thm:main}
	Let $\theta \in (0, \frac{1}{2} \pi)$ and let $\beta \in [\cos \theta, 1/\cos	\theta]$. Then using parameters $\alphaq = \beta \sqrt{(2 \log n) / d}$ and $\alphau = \sqrt{(2 \log n) / d}$ we can solve the $\theta$-ANN problem on the sphere with query/update exponents: 
	\begin{align}
	\rhoq = \left(\frac{1 - \beta \cos \theta}{\sin \theta}\right)^2 + O\left(\frac{1}{\log d}\right), \qquad \qquad \rhou = \left(\frac{\beta - \cos \theta}{\sin \theta}\right)^2 + O\left(\frac{1}{\log d}\right). \label{eq:mainp}
	\end{align}
	The resulting algorithm has a query time complexity $\tO(n^{\rhoq})$, an update time complexity $\tO(n^{\rhou})$, a preprocessing time complexity $\tO(n^{1 + \rhoq})$, and a total space complexity of $\tO(n^{1 + \rhoq})$.
	\end{theorem}
This result can equivalently be expressed in terms of $c$, by replacing $\theta = \arccos(1 - 1/c^2)$. In that case, $\theta \in (0, \frac{1}{2} \pi)$ translates to $c \in (1, \infty)$, the interval for $\beta$ becomes $\beta \in [\frac{c^2 - 1}{c^2}, \frac{c^2}{c^2 - 1}]$, and we get
\begin{align}
\rhoq = \frac{(\beta (1 - c^2) + c^2)^2}{2c^2 - 1} + O\left(\frac{1}{\log d}\right), \qquad \qquad \rhou = \frac{(1 - c^2 + \beta c^2)^2}{2c^2 - 1} + O\left(\frac{1}{\log d}\right).
\end{align}
Due to the simple dependence of these expressions on $\beta$, we can easily compute $\beta$ as a function of $\rhou$ and $\theta$ (or $c$), and substitute this expression for $\beta$ into $\rhoq$ to express $\rhoq$ in terms of $\rhou$ and $\theta$ (or $c$):
\begin{align}
\sqrt{\rhoq} = \sin \theta - \sqrt{\rhou} \cdot \cos \theta + O\left(\frac{1}{\log d}\right) = \frac{1}{c^2} \left(\sqrt{2 c^2 - 1} - \sqrt{\rhou} \cdot (c^2 - 1)\right) + O\left(\frac{1}{\log d}\right). 
\end{align}
From these expressions we can derive both Table~\ref{tab:tradeoff} and Figure~\ref{fig:tradeoff} by substituting appropriate values for $\rhoq, \rhou, \beta$ and computing Taylor expansions around $c = 1$ and $c = \infty$.

\begin{figure*}[!t]
\centering
\begin{tikzpicture}

\node at (-1cm, -1.5cm) {\Large $\cS^{d-1}$};

\begin{scope}
  \clip (1.25cm, -2.5cm) rectangle (2.5cm, 2.5cm);
  \draw[pattern=north west lines, pattern color=gray] (0, 0) circle (2.5cm);
\end{scope}

\begin{scope}
  \clip[rotate=80] ({(cos(70) * 2.5cm}, -2.5cm) rectangle (2.5cm, 2.5cm);
  \draw[pattern=north east lines, pattern color=gray] (0, 0) circle (2.5cm);
\end{scope}

\draw[black,thick] ({2.5cm * cos(60)}, {2.5cm * sin(60)}) -- ({2.5cm * cos(300)},{2.5cm * sin(300)});
\draw[black,thick] ({2.5cm * cos(150)}, {2.5cm * sin(150)}) -- ({2.5cm * cos(10)}, {2.5cm * sin(10)});

\node[black,thick] at (0,0) {\textbullet};
\node at (-0.2cm, -0.2cm) {\large $\vc{0}$};

\draw[black,dashed] (0,0) -- ({2.5cm * cos(0)}, {2.5cm * sin(0)});
\draw[black,dashed] (0,0) -- ({2.5cm * cos(80)}, {2.5cm * sin(80)});
\node at (2.8cm, 0cm) {\large $\vc{q}$};
\node at ({2.8cm * cos(80)},{2.8cm * sin(80)}) {\large $\vc{p}$};
\node[black,thick] at (2.5cm,0) {\textbullet};
\node[black,thick] at ({2.5cm * cos(80)},{2.5cm * sin(80)}) {\textbullet};

\draw[<->] (0, -0.2) -- ({2.5cm * cos(60)}, -0.2);
\node at ({1.25cm * cos(60)}, -0.45) {\large $\alphaq$};
\draw[<->] (-0.18cm, 0.04cm) -- ({cos(70) * 2.5cm * cos(80) - 0.18cm},{cos(70) * 2.5cm * sin(80) + 0.04cm});
\node at (-0.8cm, 0.5cm) {\large $\alphau$};

\draw (0.5cm,0cm) arc (0:80:0.5cm);
\node at (0.6cm, 0.4cm) {\large $\phi$};

\draw[black,thick] (0,0) circle (2.5cm);

\draw[pattern=north east lines, pattern color=gray] (4.0cm, 1.25cm) rectangle (5.3cm, 2.25cm);
\node at (7.6cm, 2cm) {$\Pr_{\vc{u} \sim \cS^{d-1}}[\vc{u} \in \text{Update}(\vc{p})]$};
\node at (6.65cm, 1.5cm) {$\propto \cC(\alphau)$};

\draw[pattern=north east lines, pattern color=gray] (4.0cm, -0.5cm) rectangle (5.3cm, 0.5cm);
\draw[pattern=north west lines, pattern color=gray] (4.0cm, -0.5cm) rectangle (5.3cm, 0.5cm);
\node at (8.6cm, 0.25cm) {$\Pr_{\vc{u} \sim \cS^{d-1}}[\vc{u} \in \text{Update}(\vc{p}) \cap \text{Query}(\vc{q})]$};
\node at (7.25cm, -0.25cm) {$\propto \cW(\alphaq, \alphau, \phi)$};

\draw[pattern=north west lines, pattern color=gray] (4.0cm, -2.25cm) rectangle (5.3cm, -1.25cm);
\node at (7.5cm, -1.5cm) {$\Pr_{\vc{u} \sim \cS^{d-1}}[\vc{u} \in \text{Query}(\vc{q})]$};
\node at (6.65cm, -2cm) {$\propto \cC(\alphaq)$};

\end{tikzpicture}
\caption{The geometry of spherical filters. A vector $\vc{p}$ is inserted into/deleted from a filter $\vc{u}$ with probability proportional to $\cC(\alphau)$, over the randomness of sampling $\vc{u}$ at random from $\cS^{d-1}$; a filter $\vc{u}$ is queried for nearest neighbors for $\vc{q}$ with probability $\cC(\alphaq)$; and a vector $\vc{p}$ at angle $\phi$ from $\vc{q}$ is found as a candidate nearest neighbor in one of the filters with probability proportional to $\cW(\alphaq, \alphau, \phi)$. \label{fig:geometry}}
\end{figure*}

\subsection{Cost analysis.} We now proceed with a proof of Theorem~\ref{thm:main}, by analyzing the costs of different steps of the filtering process in terms of the spherical cap heights $\alphaq$ and $\alphau$ and the angle of nearby vectors $\theta$, and optimizing the parameters accordingly. The analysis will be done in terms of $\theta$.

\paragraph{Updating the data structure.} The probability that a filter is considered for updates is equal to the probability that $\ip{\vc p}{\vc c} \geq \alphau$ for random $\vc{c}$, which is proportional to $\cC(\alphau)$ (cf.\ Figure~\ref{fig:geometry}). The size of $\text{Update}(\vc{p}) \subseteq C$ and the time required to compute this set with efficient decoding~\cite[Algorithm 1]{becker15lsf} are of the order $t \cdot \cC(\alphau)$. The total preprocessing time comes down to repeating this procedure $n$ times, and the total space complexity is also equal to $n \cdot t \cdot \cC(\alphau)$. (We only store non-empty buckets.)

\paragraph{Answering a query.} The probability that a filter is considered for query $\vc{q}$ is of the order $\cC(\alphaq)$ (cf.\  Figure~\ref{fig:geometry}), and the size of $\text{Query}(\vc{p})$ is of the order $t \cdot \cC(\alphaq)$. After finding the relevant buckets, we go through a number of collisions with distant vectors before (potentially) finding a near neighbor. The probability that distant vectors collide in a filter is proportional to $\cW(\alphaq, \alphau, \frac{1}{2} \pi)$ (cf.\ Figure~\ref{fig:geometry}), so the number of comparisons for all $t$ filters and all $n$ distant vectors is $\tO(n \cdot t \cdot \cW(\alphaq, \alphau, \frac{1}{2} \pi))$.

\paragraph{Choosing the number of filters.} Note that the probability that a nearby vector at angle at most $\theta$ from a query $\vc{q}$ collides with $\vc{q}$ in a random filter is proportional to $\cW(\alphaq, \alphau, \theta)$. By the union bound, the probability that two nearby vectors collide in at least one filter is at least $t \cdot \cW(\alphaq, \alphau, \theta)$. To make sure that nearby vectors are found with constant probability (say $90\%$), we set $t \propto 1 / \cW(\alphaq, \alphau, \theta)$. With the choice of $t$ fixed in terms of $\alphaq, \alphau, \theta$ and the above cost analysis in mind, the following table gives an overview of the asymptotic costs of spherical filtering in random, sparse settings.

\vspace{-0.2cm}
\begin{table}[!h]
\centering
\footnotesize
\renewcommand{\arraystretch}{1.1}
\begin{tabular}{p{8.0cm}p{7.5cm}} \toprule
\textbf{Quantity} & \textbf{Costs for general $\alphaq, \alphau, \theta$} \\ \midrule 
Time: Finding relevant filters for a query & $\cC(\alphaq) \ / \ \cW(\alphaq, \alphau, \theta)$ \\ 
Time: Comparing a query with colliding vectors & $n \cdot \cW(\alphaq,  \alphau, \frac{1}{2} \pi) \ / \ \cW(\alphaq, \alphau, \theta)$ \\ 
Time: Finding relevant filters for an update & $\cC(\alphau) \ / \ \cW(\alphaq, \alphau, \theta)$ \\
Time: Preprocessing the data & $n \cdot \cC(\alphau) \ / \ \cW(\alphaq, \alphau, \theta)$ \\
Space: Storing all filter entries & $n \cdot \cC(\alphau) \ / \ \cW(\alphaq, \alphau, \theta)$ \\ 
\bottomrule
\end{tabular}
\end{table}\vspace{-0.4cm}

\subsection{Balancing the query costs.} Next, note that answering a query consists of two steps: find the $\alphaq$-relevant filters, and go through all candidate near neighbors in these buckets. To obtain an optimal balance between these costs (with only a polynomial difference in $d$ in the time complexities), we must have $\cC(\alphaq) = d^{O(1)} \cdot n \cdot \cW(\alphaq, \alphau, \frac{1}{2} \pi)$. Raising both sides to the power $2/d$, this is equivalent to $1 - \alphau^2 = d^{O(1/d)} n^{2/d} (1 - \alphaq^2 - \alphau^2)$. Isolating $\alphau$ and noting that $n^{1/d} = \exp O(\frac{1}{\log d})$, this leads to:
\begin{align}
\alphau = d^{O(1/d)} \sqrt{\frac{n^{2/d} - 1}{n^{2/d}}} = \sqrt{\frac{2 \log n}{d}} \left(1 + O\left(\frac{1}{\log d}\right)\right).
\end{align}
This choice of $\alphau$ guarantees that the query costs are balanced. As $\alphaq$ will have a similar scaling to $\alphau$, we set $\alphaq = \beta \cdot \alphau$ for $\beta$ to be chosen later. Note that $\alphaq, \alphau = o(1)$ implies that the corresponding spherical caps (cf.\ Figure~\ref{fig:geometry}) are \textit{almost-hemispheres}, similar to spherical LSH~\cite{andoni15}. However, in our case the parameters scale as $\alphaq, \alphau = O(1/\sqrt{\log d})$, compared to $\alpha = O(1/\sqrt[4]{d})$ in~\cite{andoni15}.

\paragraph{Explicit costs.} With $\alphau$ fixed and the relation between $\alphaq$ and $\alphau$ expressed in terms of $\beta$, we now evaluate the costs for large $d$ and $n = \exp O(\frac{d}{\log d})$ in terms of $\beta$ and $\theta$. Using Taylor expansions we get:
\begin{align}
\frac{\log \cC(\alphaq)}{\log n} &= - \beta^2 + O\left(\frac{1}{\log d}\right),& \quad \frac{\log \cW(\alphaq, \alphau, \theta)}{\log n} &= - \frac{1 + \beta^2 - 2 \beta \cos \theta}{\sin^2 \theta} + O\left(\frac{1}{\log d}\right), \\
\frac{\log \cC(\alphau)}{\log n} &= -1 + O\left(\frac{1}{\log d}\right),& \quad \frac{\log \cW(\alphaq, \alphau, \frac{1}{2} \pi)}{\log n} &= - 1 - \beta^2 + O\left(\frac{1}{\log d}\right).
\end{align}
Combining these expressions, we can derive asymptotics for all of the costs related to the filtering algorithm. For the query/update exponents we then obtain 
the expressions given in Theorem~\ref{thm:main}.

\subsection{Optimal parameter range.} Note that the best complexities are obtained by choosing $\beta \in [\cos \theta, 1/\cos \theta]$; beyond this range, complexities are strictly worse. Taking inverses in this range, we get:
\begin{align}
\beta &= \frac{1 - \sqrt{\rhoq} \cdot \sin \theta}{\cos \theta} + O\left(\frac{1}{\log d}\right) = \cos \theta + \sqrt{\rhou} \cdot \sin \theta + O\left(\frac{1}{\log d}\right). \label{eq:beta}
\end{align}
Isolating $\sqrt{\rhoq}$ then leads to~\eqref{eq:main}, while \eqref{eq:beta} also shows how to choose $\beta$ to achieve given complexities. 


\section{Asymmetric spherical filters for dense regimes} 
\label{sec:dense}

We now revisit the dense regime of data sets of size $n = 2^{\Theta(d)}$, as previously analyzed in~\cite{becker15lsf} for symmetric ANN. We will again use two parameters $\alphaq, \alphau \in [0,1]$ where the optimization now leads to a slightly different, more refined result, depending on the chosen tradeoff.

\subsection{Random dense instances.} To study the dense regime, we consider the following model.

\begin{definition}[Random $\theta$-ANN in the dense regime]
Given an angle $\theta \in (0, \frac{1}{2} \pi)$, a query $\vc{q} \in \R^d$, and a data set $\cD$ of $n = 2^{\Theta(d)}$ points sampled uniformly at random from $\cS^{d-1}$, the random $\theta$-ANN problem is defined as the problem of either finding a point $\vc{p} \in \cD$ with $\phi(\vc{p}, \vc{q}) \leq \theta$, or concluding that with high probability no vector $\vc{p} \in \cD$ exists with $\phi(\vc{p}, \vc{q}) \leq \theta$.
\end{definition}

At first sight, the above definition does not seem to correspond to an \textit{approximate}, but to an \textit{exact} nearest neighbor instance. However, we made a critical assumption on $\cD$ here: we assume that these points are sampled uniformly at random from the sphere. This seems to be a natural assumption in various applications (see e.g.\ \cite{herold15, laarhoven15lsh, may15}), and this implies that in fact many of the points in the data set lie at angle approximately $\frac{1}{2} \pi$ from $\vc{q}$. As a result the problem is significantly easier than e.g.\ the worst-case ANN setting with $c \approx 1$, where the entire data set might lie at angle $\theta + \delta$ from $\vc{q}$. 

For comparing this problem with the sparse setting of Section~\ref{sec:sparse}, observe that this dense problem is \textit{harder}: uniformly random points on the sphere (of which roughly half has angle less than $\frac{1}{2} \pi$ with $\vc{q}$) are more likely to cause collisions than orthogonal points to $\vc q$. The number of collisions with distant vectors will therefore increase, and we expect the query and update exponents to be larger. This was also observed in e.g.\ \cite{becker15lsf, becker15cp, laarhoven15lsh, laarhoven15sphere}, where the exponent for lattice sieving with other ANN techniques would have been smaller if one could only assume that far away means orthogonal.

Note that we could also extend the analysis of Section~\ref{sec:sparse} to the dense regime simply by fixing the distant angle at $\psi = \frac{\pi}{2}$. In that case, similar to~\cite{becker15lsf}, both the query and update exponents will become smaller compared to the low-density regime as the problem becomes easier. However, such an assumption would imply that the data set is not spherically symmetric and is concentrated on one part of the sphere, in which case data-dependent methods may be preferred~\cite{leskovec14, wang14}.

\subsection{Density reduction and the critical density.} As $\cD$ is sampled at random from $\cS^{d-1}$, a point $\vc{p} \in \cD$ is close to $\vc{q}$ with probability proportional to $(\sin \theta)^d$. With $n$ points, we expect to find approximately $n \cdot (\sin \theta)^d$ nearby neighbors $\vc{p} \in \cD$. For $n \ll (\sin \theta)^{-d}$, nearby vectors are rare, and we are essentially solving the exact (decisional) nearest neighbor problem with high probability. On the other hand, if $n \gg (\sin \theta)^{-d}$, then we expect there to be many ($n^{O(1)}$) solutions $\vc{p} \in \cD$ with $\phi(\vc{p}, \vc{q}) \leq \theta$. 

In our analysis we will focus on the case where $n = \tO((\sin \theta)^{-d})$: there might not be any near neighbors in $\cD$ at all, but if there is one, we want to find it. For the regime $n \gg (\sin \theta)^{-d}$, we can reduce this problem to a regime with a lower density $n' = \tO((\sin \theta)^{-d})$ through a simple transformation:
\vspace{-0.1cm}
\begin{itemize}
	\item Randomly select a subset $\cD' \subset \cD$ of size $n' = \tO((\sin \theta)^{-d})$.
	\itemsep-.3em \item Run the (approximate) nearest neighbor algorithm on this subset $\cD'$.
\end{itemize}
By choosing the hidden factor inside $n'$ sufficiently large, with high probability there will be a solution in this smaller subset $\cD'$ as well, which our algorithm will find with high probability. This means that in our cost analysis, we can then simply replace $n$ by $n'$ to obtain the asymptotic complexities after this density reduction step. We denote the regime $n \propto (\sin \theta)^{-d}$ as the \textit{critical density}.

Note that if we are given a random data set of size $n$, then we expect the nearest neighbor to a random query $\vc{q} \in \cS^{d-1}$ to lie at angle $\theta \approx \arcsin(n^{-1/d})$ from $\vc{q}$; for larger angles we will find many solutions, while for smaller angles w.h.p.\ there are no solutions at all (except for \textit{planted} near neighbor instances (\textit{outliers}) which are not random on the sphere). This further motivates why the critical density is important, as $\theta \approx \arcsin(n^{-1/d})$ commonly corresponds to solving the \textit{exact} nearest neighbor problem for random data sets. Setting $\theta \ll \arcsin(n^{-1/d})$ then corresponds to searching for outliers. 

\subsection{Main result.} We first state the main result for the random, dense setting described above without making any assumptions on the density. A derivation of Theorem~\ref{thm:maindense} can be found in Appendix~\ref{sec:app}.
	\begin{theorem} \label{thm:maindense}
	Let $\theta \in (0, \frac{1}{2} \pi)$ and let $\beta \in [\cos \theta, 1 / \cos \theta]$. Then using parameters $\alphaq = \beta \sqrt{1 - n^{-2/d}}$ and $\alphau = \sqrt{1 - n^{-2/d}}$ we can solve the dense $\theta$-ANN problem on the sphere with exponents: 
	\begin{align}
	\rhoq &= \frac{-d}{2 \log n} \log \left[1 - \left(1 - n^{-2/d}\right) \frac{1 + \beta^2 - 2 \beta \cos \theta}{\sin^2 \theta}\right] + \frac{d}{2 \log n} \log \left[1 - \left(1 - n^{-2/d}\right) \beta^2 \right], \\ 
	\rhou &= \frac{-d}{2 \log n} \log \left[1 - \left(1 - n^{-2/d}\right) \frac{1 + \beta^2 - 2 \beta \cos \theta}{\sin^2 \theta}\right] - 1. \label{eq:main2}
	\end{align}
	\end{theorem}
Note that in the limit of $n^{1/d} \to 1$, corresponding to sparse data sets, we obtain the expressions from Theorem~\ref{thm:main}. This matches our intuition that taking $n$ points uniformly at random from the sphere with $n = 2^{o(d)}$ roughly means that all points have angle $\psi = \frac{1}{2} \pi$ with a query $\vc{q}$, as described in Section~\ref{sec:sparse-random}. 

\subsection{Critical densities.} As a special case of the above result, we focus on the regime of $n \propto (\sin \theta)^{-d}$. The following result shows the complexities obtained after substituting this density into Theorem~\ref{thm:maindense}.
	\begin{corollary} \label{thm:main3}
	Let $\theta \in (0, \frac{1}{2} \pi)$, let $\beta \in [\cos \theta, 1/\cos \theta]$, and let $n = (1 / \sin \theta)^d$. Then using parameters $\alphaq = \beta \cos \theta$ and $\alphau = \cos \theta$, the complexities in Theorem~\ref{thm:maindense} reduce to: 
	\begin{align}
	n^{\rhoq} &= \left(\frac{\sin^2 \theta \, (\beta \cos \theta + 1)}{\beta  \cos \theta - \cos 2 \theta}\right)^{d/2}, \qquad n^{\rhou} = \left(\frac{\sin^2 \theta}{1 - \cot^2 \theta\left(\beta^2 - 2 \beta \cos \theta + 1\right)}\right)^{d/2}. \label{eq:main3}
	\end{align}
	\end{corollary}
To obtain further intuition into the above results, let us consider the limits obtained by setting $\beta \in \{\cos \theta, 1, 1/\cos \theta\}$. For $\beta = \cos \theta$, we obtain exponents of the order:
\begin{align}
n \propto (\sin \theta)^{-d}, \quad \beta = \cos \theta, \qquad \implies \qquad \rhou = 0, \quad \rhoq = \dfrac{\log 2 - \log(3 + 2 \cos (2 \, \theta))}{2 \log (\sin \theta)}\, . \kern 0.8cm
\end{align}
Balancing the complexities is done by setting $\beta = 1$, in which case we obtain the asymptotic expressions:
\begin{align}
n \propto (\sin \theta)^{-d}, \quad \beta = 1, \qquad \implies \qquad \rhoq = \rhou = \frac{2 \log(\tan \frac{\theta}{2}) + \log(2 \cos \theta + 1)}{2 \log(\sin \theta)} - 1.
\end{align}
To minimize the query complexity, we would ideally set $\beta = 1 / \cos \theta$, as then the query exponent approaches $0$ for large $n$. However, one can easily verify that substituting $\beta = 1 / \cos \theta$ into~\eqref{eq:main3} leads to a denominator of $0$, i.e.\ the update costs and the space complexities blow up as $\beta$ approaches $1 / \cos \theta$. To get an idea of how the update complexity scales in terms of the query complexity, we set $\rhoq = \delta$ and we compute a Taylor expansion around $\delta = 0$ for $\rhou$ to obtain:
\begin{align}
n \propto (\sin \theta)^{-d}, \quad \beta = 1 \qquad \implies \qquad \rhoq = \delta, \quad \rhou = \frac{\log \left(8 \, \delta \log(1 / \sin \theta) \tan^2 \theta\right)}{2 \log(\sin \theta)} - 1 + O(\delta).
\end{align}
In other words, for fixed angles $\theta$, to achieve $\rhoq = \delta$ the parameter $\rhou$ scales as $\log(1 / \delta)$. Note that except for the latter result, we can substitute $\cos \theta = 1 - 1 / c^2$ and $c = 1 + \eps$, and compute a Taylor series expansion of these expressions around $\eps = 0$ to obtain the expressions in Table~\ref{tab:tradeoff} for small $c = 1 + \eps$. This matches our intuition that $\theta \to \frac{1}{2} \pi$ for random data sets corresponds to $\theta \approx \frac{1}{2} \pi$ for sparse settings.

Finally, we observe that substituting $\theta = \frac{1}{3} \pi$, for a minimum space complexity we obtain $\rhoq = \log(\frac{5}{4}) / \log(\frac{4}{3})$, while balancing both costs leads to $\rhoq = \rhou = \log(\frac{9}{8}) / \log(\frac{4}{3})$. These results match those derived in~\cite{becker15lsf} for the application to sieving for finding shortest lattice vectors.


\section{Discussion and open problems} 
\label{sec:discussion}

We conclude this work with a brief discussion on how the described methods can possibly be extended and modified to solve other problems and to obtain a better performance in practical applications.

\paragraph{Probing sequences.} In LSH, a common technique to reduce the space complexity, or to reduce the number of false negatives, is to use probing~\cite{andoni15cp, lv07, panigrahy06}: one does not only check \textit{exact} matches in the hash tables for reductions, but also \textit{approximate} matches which are still more likely to contain near neighbors than random buckets. Efficiently being able to define a \textit{probing sequence} of all buckets, in order of likelihood of containing near neighbors, can be useful both in theory and in practice.

For spherical filters, an optimal probing sequence is obtained by sorting the filters (code words) according to their inner products with the target vector. Due to the large number of buckets $t$, computing and sorting all filter buckets is too costly, but for practical applications we can do the following. We first choose a sequence $1 = \alpha_0 > \alpha_1 > \dots > \alpha_T$, and then given a target $\vc{t}$ we apply our decoding algorithm to find all code words $\vc{c} \in C$ with $\ip{\vc{c}}{\vc{t}} \in (\alpha_1, \alpha_0]$. The corresponding buckets are the most likely to contain nearby vectors. If this does not result in a nearest neighbor, we apply our decoding algorithm to find code words $\vc{c} \in C$ with $\ip{\vc{c}}{\vc{t}} \in (\alpha_2, \alpha_1]$, and we repeat the above procedure until e.g.\ we are convinced that no solution exists. For constant $T$, the overhead of this repeated decoding is small.

To implement the search for finding code words $\vc{c} \in C$ with $\ip{\vc{c}}{\vc{t}} \in (\alphal, \alphau]$ efficiently, we can use Algorithm~\ref{alg:interval} in Appendix~\ref{sec:appb}. The most costly step of this decoding algorithm is computing and sorting all blockwise inner products $\ip{\vec c_{k,j}}{\vec t_k}$, but note that these computations have to be performed only once; later calls to this function with different intervals $(\alpha_{i+1}, \alpha_i]$ can reuse these sorted lists. 

\paragraph{Ruling out false negatives.} An alternative to using probing sequences to make sure that there are no false negatives, is to construct a scheme which guarantees that there are never any false negatives at all (see e.g.\ \cite{pagh16}). In the filtering framework, this corresponds to using codes $C$ such that it is guaranteed that nearby vectors always collide in one of the filters. In other words, for each pair of points $\vec p, \vec q$ on the sphere at angle $\theta$, the corresponding wedge $\cW_{\vec p, \alphau, \vec q, \alphaq}$ must contain a code word $\vc{c} \in C$. Note that with our random construction we can only show that with high probability, this is the case.

For spherical filters, codes guaranteeing this property correspond to spherical codes such that all possible wedges $\cW_{\vec p, \alphau, \vec q, \alphaq}$ for $\vc{p}, \vc{q} \in \cS^{d-1}$ contain at least one code word $\vc{c} \in C$. For $\alphaq = \alphau = \alpha$, note that at the middle of such a wedge lies a point $\vec y = (\vec p + \vec q) / \|\vec p + \vec q\|$ at angle $\theta/2$ from both $\vec p$ and $\vec q$. If a code is not covering and allows for false negatives, then there are no code words at angle $\frac{1}{2} \theta - \arccos \alpha$ from $\vec y$. In particular, the \textit{covering radius} of the code (the smallest angle $\psi$ such that spheres of angle $\psi$ around all code words cover the entire sphere) is therefore larger than $\frac{1}{2} \theta - \arccos \alpha$. Equivalently, being able to construct spherical codes of low cardinality with covering radius at most $\frac{1}{2} \theta - \arccos \alpha$ implies being able to construct a spherical filtering scheme without false negatives.

As we make crucial use of concatenated codes $C = C_1 \times \dots \times C_m$ to allow for efficient decoding, covering codes without efficient decoding algorithms cannot be used for $C$. Instead, one might aim at using such covering codes for the subcodes: if all subcodes $C_i$ have covering radius at most $\frac{1}{2} \theta - \arccos \frac{\alpha}{m}$, then the concatenated code $C = C_1 \times \dots \times C_m$ has a covering radius of at most $\frac{1}{2} \theta - \arccos \alpha$. Finding tight bounds on the size of a spherical code with covering radius $\frac{1}{2} \theta - \arccos \frac{\alpha}{m}$ (where $\theta$ is defined by the problem setting, and $\alpha$ may be chosen later) would directly lead to an upper bound on the number of filters needed to guarantee that there are no false positives.

\paragraph{Sparse codes for efficiency.} As described in e.g.\ \cite{achlioptas01, becker15lsf, li06}, it is sometimes possible to use sparse, rather than fully random codes without losing on the performance of a nearest neighbor scheme. Using sparse subcodes $C_i$ might further reduce the overhead of decoding (computing blockwise inner products). For this we could either use randomly sampled sparse subcodes, but one might also consider using codes which are guaranteed to be ``smooth'' on the sphere and have a small covering radius. Similar to Leech lattice LSH~\cite{andoni06}, one might consider using vectors from the Leech lattice in $24$ dimensions to define the subcodes. Asymptotically in our construction the block size $d/m$ needs to scale with $d$, and fixing $d/m = 24$ would invalidate the proof of smoothness of~\cite[Theorem 5.1]{becker15lsf}, but in practice both $n$ and $d$ are fixed, and only practical assessments can show whether predetermined spherical codes can be used to obtain an even better performance.

\paragraph{Optimality of the tradeoff.} An interesting open problem for future work is determining precise bounds on the best tradeoff that one can possibly hope to achieve in the sparse regime. Since our tradeoff matches various known bounds in the regimes of small and large approximation factors $c$~\cite{andoni06b, andoni09, kushilevitz00, panigrahy08, panigrahy10} and no LSH scheme can improve upon our results in the balanced setting~\cite{andoni15lb, dubiner10}, and since the tradeoff can be described through a remarkably simple relation (especially when described in terms of $\theta$), we conjecture that this tradeoff is optimal.
\begin{conjecture}
Any algorithm for sparse $\theta$-ANN on the sphere must satisfy $\sqrt{\rhoq} + \sqrt{\rhou} \cdot \cos \theta \geq \sin \theta$.
\end{conjecture}
As a first step, one might try to (dis)prove this conjecture within the LSH framework, similar to various other works focusing on lower bounds for schemes that fall in this category~\cite{motwani07, odonnell14}.

\paragraph{Extension to Euclidean spaces.} As mentioned in the introduction, Andoni and Razenshteyn showed how to reduce ANN in Euclidean spaces to (sparse) random ANN on the sphere in the symmetric case, using LSH techniques~\cite{andoni15}. An important open problem for future work is to see whether the techniques and the reduction described in~\cite{andoni15} are compatible with locality-sensitive filters, and with asymmetric nearest neighbor techniques such as those presented in this paper. If this is possible, then our results may also be applicable to all of $\ell_2^d$, rather than only to the angular distance on $\R^d$ or to Euclidean distances on the unit sphere $\cS^{d-1}$. 

\paragraph{Combination with cross-polytope LSH.} Finally, the recent paper~\cite{andoni15cp} showed how cross-polytope hashing (previously introduced by Terasawa and Tanaka~\cite{terasawa07}) is asymptotically equally suitable for solving Euclidean nearest neighbor problems on the sphere (and for the angular distance) as the approach of spherical LSH of using large, completely random codes on the sphere~\cite{andoni15}. Advantages of cross-polytope LSH over spherical LSH are that they have a much smaller size (allowing for faster decoding), and that cross-polytope hash functions can be efficiently rerandomized using sparse and fast random projections such as Fast Hadamard Transforms~\cite{andoni15cp}. In that sense, cross-polytope LSH offers a significant practical improvement over spherical LSH.

The approach of using spherical filters is very similar to spherical LSH: large, random (sub)codes are used to define regions on the sphere. A natural question is therefore whether ideas analogous to cross-polytope hashing can be used in combination with spherical filters, to reduce the subexponential overhead in $d$ for decoding to an overhead which is only polynomial in $d$. This is also left as an open problem for further research.


\newcommand{\etalchar}[1]{$^{#1}$}
\providecommand{\noopsort}[1]{}

\appendix


\section{Tradeoff figure in the sparse regime}
\label{sec:fig}

\begin{figure*}[!t]
\centering
\includegraphics[width=\linewidth]{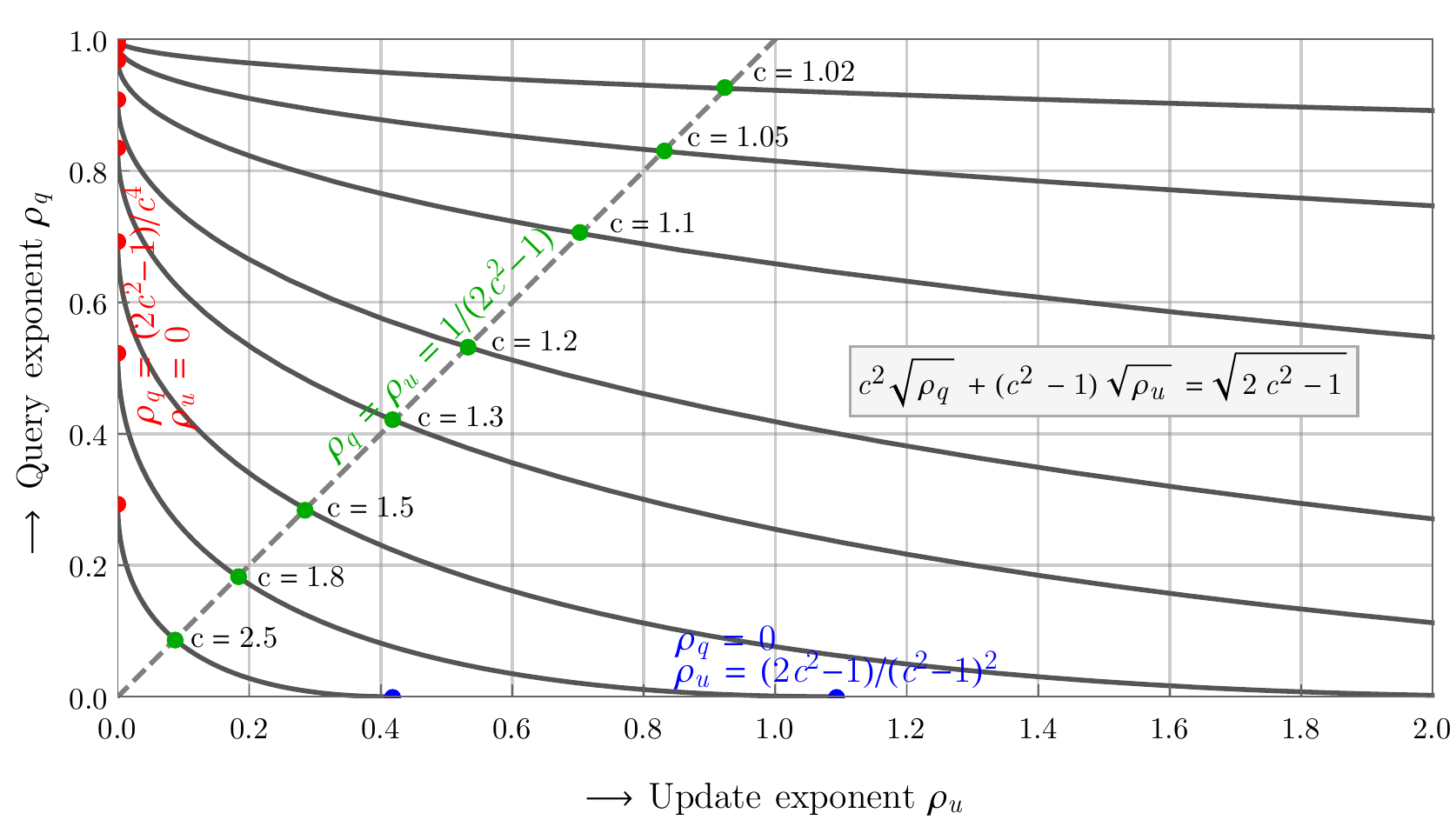}
\caption{Tradeoffs between the query and update complexities, for various $c$. Informally, the $x$-axis represents the space and the $y$-axis the time (for queries). The diagonal represents symmetric ANN. \label{fig:tradeoff}}
\end{figure*}

Figure~\ref{fig:tradeoff} describes asymptotic tradeoffs for different values of $c$. Note that the query exponent is always smaller than $1$, regardless of $c > 1$, and note that the update exponent is smaller than $1$ (and the query exponent less than $1/2$) for all $c > \sqrt{2 + \sqrt{2}} \approx 1.85$ corresponding to $\theta = \frac{1}{4} \pi$.

\section{Analysis for dense regimes}
\label{sec:app}

To derive the complexities for the dense regime of $n = 2^{\Theta(d)}$, we follow the same approach as for the sparse regime of $n = 2^{o(d)}$, but without making the assumption that e.g.\ $n^{1/d} = 1 + o(1)$. 

\subsection{General expressions.} First, we observe that the cost analysis described in Section~\ref{sec:sparse} applies in the dense setting as well, with the modification that we no longer assume that $\vc{q}$ is orthogonal to all of $\cD$ (except for a potential nearest neighbor). The update and query costs remain the same as before in terms of $t$ and the volumes $\cC(\cdot)$ and $\cW(\cdot)$, but to obtain the number of collisions with distant vectors, we take a different angle. First, observe that each vector is added to $\tO(t \cdot \cC(\alphau))$ filters, and that we have $n$ vectors, leading to $\tO(n \cdot t \cdot \cC(\alphau))$ total entries in the filters or $\tO(n \cdot t \cdot \cC(\alphau))$ entries in each filter\footnote{Here a crucial property of the filters is that each filter covers a region of exactly the same size on the sphere.}. For a given vector $\vc{q}$, we then query $\tO(t \cdot \cC(\alphaq))$ buckets for nearest neighbors. In total, we therefore expect to find $\tO(n \cdot t \cdot \cC(\alphau) \cdot \cC(\alphaq))$ colliding vectors for a query vector. Again setting $t \propto 1 / \cW(\alphaq, \alphau, \theta)$ to make sure that nearby vectors are found with constant probability, we obtain the following updated table of the asymptotic costs of spherical filtering in random, dense settings.

\begin{table}[!h]
\centering
\footnotesize
\renewcommand{\arraystretch}{1.1}
\begin{tabular}{p{8.0cm}p{7.5cm}} \toprule
\textbf{Quantity} & \textbf{Costs for $\alphaq, \alphau, \theta$} \\ \midrule 
Time: Finding relevant filters for a query & $\cC(\alphaq) \ / \ \cW(\alphaq, \alphau, \theta)$ \\ 
Time: Comparing a query with colliding vectors & \textcolor{blue}{$n \cdot \cC(\alphaq) \cdot \cC(\alphau) \ / \ \cW(\alphaq, \alphau, \theta)$} \\ 
Time: Finding relevant filters for an update & $\cC(\alphau) \ / \ \cW(\alphaq, \alphau, \theta)$ \\
Time: Preprocessing the data & $n \cdot \cC(\alphau) \ / \ \cW(\alphaq, \alphau, \theta)$ \\
Space: Storing all filter entries & $n \cdot \cC(\alphau) \ / \ \cW(\alphaq, \alphau, \theta)$ \\ 
\bottomrule
\end{tabular}
\end{table}\vspace{-0.4cm}

\subsection{Balancing the query costs.} Next, to make sure that the query costs are balanced, and not much more time is spent on looking for relevant filters rather than actually doing comparisons, we again look for parameters such that these costs are balanced. In this case we want to solve the asymptotic equation $\cC(\alphaq) = n \cdot \cC(\alphaq) \cdot \cC(\alphau)$ or $\cC(\alphau) = (1 - \alphau^2)^{d/2} = 1/n$. Solving for $\alphau$ leads to $\alphau = \sqrt{1 - n^{-2/d}}$ leading to the parameter choice described in Theorem~\ref{thm:maindense}. For now we again set $\alphaq = \beta \cdot \alphau$ with $\beta$ to be chosen later.

\subsection{Explicit costs.} We now evaluate the costs for large $d$ and $n = 2^{\Theta(d)}$, in terms of the ratio $\beta$ between the two parameters $\alphaq$ and $\alphau$, and the nearby angle $\theta$. This leads to $\cC(\alphau) = 1/n$ and:
\begin{align}
\cC(\alphaq) = \left(1 - (1 - n^{-2/d}) \, \beta^2\right)^{d/2}, \ \quad \cW(\alphaq, \alphau, \theta) = \left(1 - (1 - n^{-2/d}) \, \frac{1 + \beta^2 - 2 \beta \cos \theta}{\sin^2 \theta}\right)^{d/2}.
\end{align}
Combining these expressions, we can then derive asymptotic estimates for all of the costs of the algorithm. For the query and update exponents $\rhoq = \log[\cC(\alphaq) / \cW(\alphaq, \alphau, \theta)] / \log n$ and $\rhou = \log[\cC(\alphau) / \cW(\alphaq, \alphau, \theta)] / \log n$ we then obtain:
\begin{align}
\rhoq &= \frac{-d}{2 \log n} \log \left[1 - \left(1 - n^{-2/d}\right) \frac{1 + \beta^2 - 2 \beta \cos \theta}{\sin^2 \theta}\right] + \frac{d}{2 \log n} \log \left[1 - \left(1 - n^{-2/d}\right) \beta^2 \right], \\ 
\rhou &= \frac{-d}{2 \log n} \log \left[1 - \left(1 - n^{-2/d}\right) \frac{1 + \beta^2 - 2 \beta \cos \theta}{\sin^2 \theta}\right] - 1.
\end{align}
These are also the expressions given in Theorem~\ref{thm:maindense}.

\subsection{Optimal parameter range.} We observe that again the best exponents $\rhoq$ and $\rhou$ are obtained by choosing $\beta \in [\cos \theta, 1/\cos \theta]$; beyond this range, the complexities are strictly worse. This completes the derivation of Theorem~\ref{thm:maindense}.


\section{Interval decoding}
\label{sec:appb}

Algorithm~\ref{alg:interval} describes how to perform list-decoding for intervals, which may be relevant in practice for e.g.\ computing probing sequences as described in Section~\ref{sec:discussion}. The algorithm is based on \cite[Algorithm 1]{becker15lsf}, where now two sets of bounds are maintained to make sure that we only consider solutions which lie within the given range, rather than above a threshold. The bounds $L_k$ and $U_k$ indicate the minimum and maximum sum of inner products that can still be obtained in the last $m - k$ sorted lists of vectors and inner products; if in the nested for-loops, the current sum of inner products $\sum_i d_{i, j_i}$ is not in the interval $(L_k, U_k]$, then there are no solutions anymore in the remaining part of the tree. Conversely, if this sum of inner products does lie in the interval, then there must be at least one solution.

\begin{algorithm}[!t]
\small
\caption{EfficientIntervalDecoding$(C, \vec t, \alphal, \alphah)$}
\label{alg:interval}
\begin{algorithmic}[1]
\Require The description $C_1, \dots, C_m$ of the code $C$; a target vector $\vec{t} \in \R^d$; and $0 \leq \alphal < \alphah \leq 1$.
\Ensure Return all code words $\vc{c} \in C$ with $\ip{\vc t}{\vc{c}} \in (\alphal, \alphah]$
\State Sort each list $C_k$ by decreasing dot-products $d_{k,j} = \ip{\vec t_k}{\vec c_{k,j}}$ with $\vec{t}_k$. 
\State Precompute $m$ bounds $L_k = \alphal - \sum_{i = k + 1}^m d_{i, t^{1/m}}$. 
\State Precompute $m$ bounds $U_k = \alphah - \sum_{i = k + 1}^m d_{i, 1}$.
\State Initialize an empty output set $S\leftarrow \emptyset$.
\State Compute the lower bound $\ell_1 = \min\{j_1: d_{1, j_1} > L_1\}$.
\Comment{do a binary search over $[1, t^{1/m}]$}
\State Compute the upper bound $u_1 = \max\{j_1: d_{1, j_1} \leq U_1\}$.
\Comment{do a binary search over $[1, t^{1/m}]$}
\For{\textbf{each} $j_1 \in \{\ell_1, \dots, u_1\}$}
\State Compute the lower bound $\ell_2 = \min\{j_2: d_{2, j_2} > L_2 - d_{1, j_1}\}$.
\State Compute the upper bound $u_2 = \max\{j_2: d_{2, j_2} \leq U_2 - d_{1, j_1}\}$.
\For{\textbf{each} $j_2 \in \{\ell_2, \dots, u_2\}$}
\State [...]
\State Compute the lower bound $\ell_m = \min\{j_m: d_{m, j_m} > L_m - \sum_{k=1}^{m-1} d_{k, j_k}\}$.
\State Compute the upper bound $u_m = \max\{j_m: d_{m, j_m} \leq U_m - \sum_{k=1}^{m-1} d_{k, j_k}\}$.
\For{\textbf{each} $j_m \in \{\ell_m, \dots, u_m\}$}
\State Add the code word $\vec c = (\vec{c}_{1,j_1},\dots,\vec{c}_{m,j_m})$ to $S$.
\EndFor
\State [...]
\EndFor
\EndFor
\State \Return $S$
\end{algorithmic}
\end{algorithm}

\end{document}